\documentclass[a4paper,11pt]{article}
\usepackage{pos}

\title{Chiral susceptibility and axial U(1) anomaly near the (pseudo-)critical temperature}

\author[a,b]{JLQCD Collaboration: S.~Aoki}
\author[c]{Y.~Aoki}
\author*[d]{H.~Fukaya}
\author[e,f]{S.~Hashimoto}
\author[c]{I.~Kanamori}
\author[e,f,g]{T.~Kaneko}
\author[c]{Y.~Nakamura}
\author[h]{K.~Suzuki}
\author[d]{D.~Ward}

\affiliation[a]{
  Center for Gravitational Physics, Yukawa Institute for Theoretical Physics,
        Kyoto University, Kyoto 606-8502, Japan}

\affiliation[b]{RIKEN Nishina Center (RNC), Saitama 351-0198, Japan}

\affiliation[c]{RIKEN Center for Computational Science, 7-1-26
Minatojima-minami-machi, Chuo-ku, Kobe, Hyogo 650-0047, Japan}

\affiliation[d]{Department of Physics, Osaka University, 
        Toyonaka, Osaka 560-0043 Japan}

\affiliation[e]{High Energy Accelerator Research Organization (KEK),
        Tsukuba 305-0801, Japan}
\affiliation[f]{School of High Energy Accelerator Science,
        The Graduate University for Advanced Studies
        (Sokendai),Tsukuba 305-0801, Japan}
\affiliation[g]{Kobayashi-Maskawa Institute for the Origin of Particles and the Universe, Nagoya University,
Aichi 464-8603, Japan}

\affiliation[h]{Advanced Science Research Center, Japan Atomic
  Energy Agency (JAEA), Tokai 319-1195, Japan
  }

\emailAdd{hfukaya@het.phys.sci.osaka-u.ac.jp}

\abstract{
We investigate relations between the chiral susceptibility and axial $U(1)$ anomaly in lattice QCD at high temperatures. Employing the exactly chiral symmetric Dirac operator, we separate the purely axial $U(1)$ breaking effect in the connected and disconnected chiral susceptibilites in a theoretically clean manner. Preliminary results for 
two-flavor lattice QCD near the critical temperature are presented.
\par
Preprint number: OU-HET-1211, KEK-CP-0395
}

\FullConference{The 40th International Symposium on Lattice Field Theory (Lattice 2023)\\
July 31st - August 4th, 2023\\
Fermi National Accelerator Laboratory\\}

\renewcommand{\logo}{\relax}

\begin{document}
\maketitle

\section{Introduction}

In the current universe, the chiral $SU(2)_L\times SU(2)_R$ symmetry
is spontaneously broken, which explains the origin of the hadron masses
as well as the special role of the pion as the (pseudo) Nambu-Goldstone boson.
It is widely believed that at the critical temperature $T_c\sim 150$ MeV, 
the chiral symmetry is recovered.
This is the so-called chiral phase transition
our universe experienced just 10 $\mu$s after the Big-Bang.
The chiral condensate $\langle \bar{q}q\rangle$
is considered to be 
the order parameter of the phase transition of QCD.

The QCD partition function at temperature $T$
is given by a functional
integral over the gluon field $A$,
\begin{align}
Z(m,T) = \int [dA] \det(D(A,T)+m)^{N_f}e^{-S_G(A,T)}, 
\end{align}
where $D(A,T)$ denotes the Dirac operator of quarks
and $S_G(A,T)$ is the Yang-Mills acion of gauge fields.
The $T$ dependence is encoded as a finite temporal direction size
$L_t=1/T$ with the periodic boundary condition for gluons
and anti-periodic one for quarks.
We take $N_f=2$ with degenerate up and down quark masses
$m_{u}=m_{d}=m$ and take the strange and heavier quarks
just as spectators.
The chiral condensate is equivalent to
the first derivative of $Z(m,T)$ with respect to $m$,
\begin{align}
\Sigma(m,T)=-\langle \bar{q}q\rangle_{m,T}
  = \frac{1}{N_fV}\frac{\partial}{\partial m}\ln Z(m,T),
\end{align}
where $V$ is the volume of the Euclidean spacetime.
In this work, we focus on the second derivative,
\begin{align}
\chi(m,T)
  = \frac{\partial}{\partial m}\Sigma(m,T),
\end{align}  
which is known as the chiral susceptibility (See recent developments in 
\cite{Bhattacharya:2014ara, Bonati:2015bha, Brandt:2016daq, Taniguchi:2016ofw, Ding:2019prx}).

\begin{figure*}[bth]
  \centering
    \includegraphics[width=14cm]{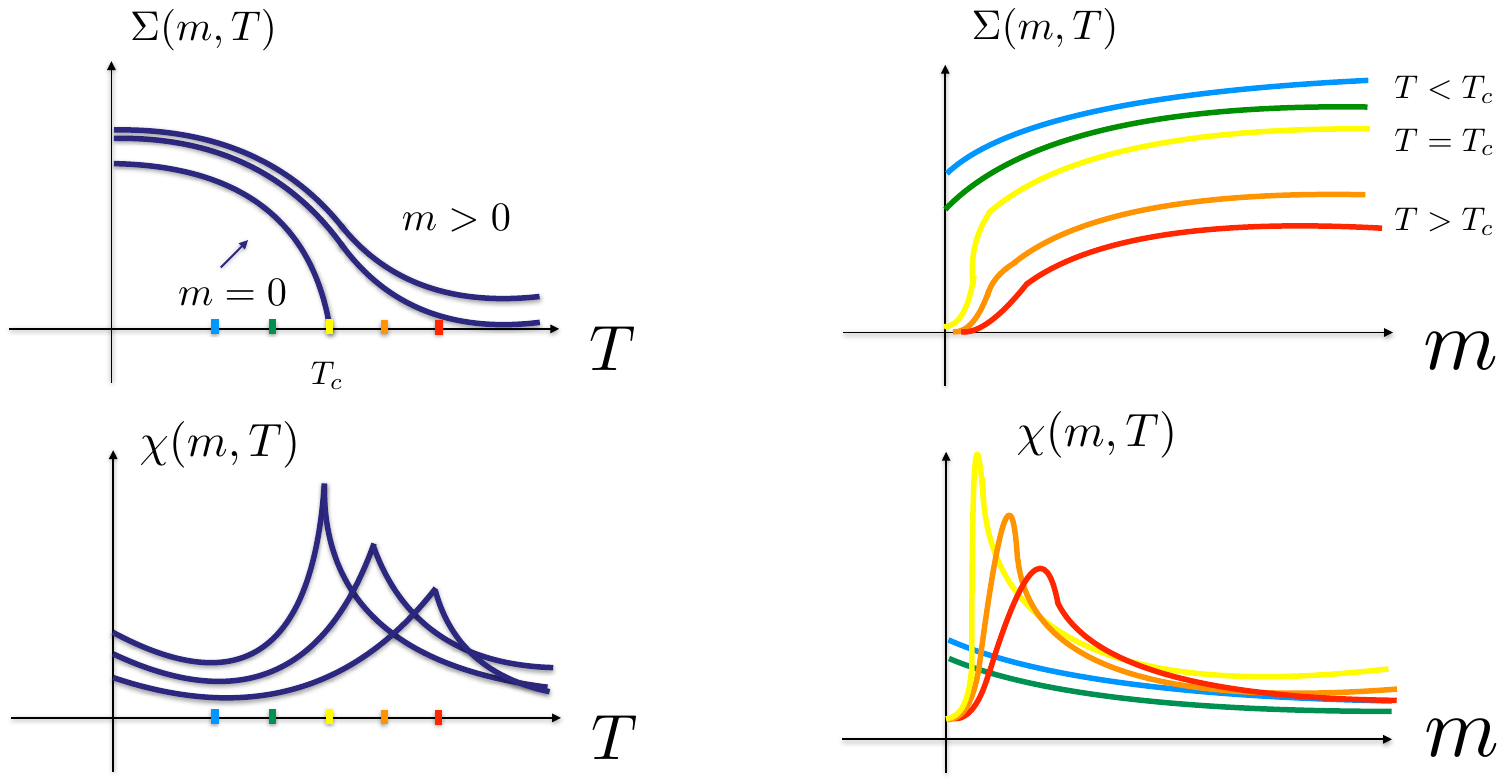}
  \caption{
    Schematic picture of the 
    chiral condensate (top panels) and susceitpibility (bottom).
    The temperature $T$ dependence (left panels) and 
    and that on the quark mass $m$ (right) at fixed temperatures
    marked by colored symbols are shown.
  }
  \label{fig:Tmdep}
\end{figure*}

In Fig.~\ref{fig:Tmdep} we illustrate how the chiral condensate
and susceptiblity depend on $T$ and $m$
in the typical scinario of the second-order
phase transition.
At $m=0$, the condensate continuously disappears at $T=T_c$,
while it becomes a crossover for $m>0$.
It is, therefore, important to trace the quark mass
dependence of the condensate at fixed temperatures
marked by the color symbols at the top-left panel.
As the top-right panel shows, while a rather mild quark mass
dependence to a nonzero value at $m=0$ is
expected for lower temperatures than $T_c$, while
the higher $T$ curves show a steep drop to zero
near the chiral limit $m=0$. This drop should be
reflected to the peaks of the chiral susceptibility $\chi(m,T)$
as the bottom-right panel presents.

When the phase transition is the first order, 
the $T$ dependence becomes discontinuous at $T=T_c$
but in lattice QCD in a finite volume
the discontinuity is smoothed and 
the situation will not be much different from Fig.~\ref{fig:Tmdep}.

The chiral condensate is a probe of the $SU(2)_L\times SU(2)_R$ 
chiral symmetry breaking or restoration. 
But the operator condensate $\langle \bar{q} q\rangle $ also breaks 
the axial $U(1)$ symmetry or $U(1)_A$ in short. 
In this talk we would like to discuss how much the $U(1)_A$  
part contributes to the phase transition.

Since the anomaly is an explicit breaking in the theory 
given at the cutoff scale,
it is natural to assume that the axial $U(1)$ 
is broken at any energy scale, and it is
insensitive to the infra-red scales $m$ and $T$.
If this is true, the $m$ and $T$ dependence of the
chiral condensate or chiral susceptibility should
reflect the $SU(2)_L\times SU(2)_R$ breaking/restoration rather
than that of $U(1)_A$.

A different scinario is possible if the topological excitations
or instantons of the gluons are responsible for the low energy dynamics of QCD.
In a paper by Callan, Dashen and Gross \cite{Callan:1977gz}, 
they computed how much
the instantons enhances the $U(1)_A$ anomaly as
a trigger for the spontaneous $SU(2)_L\times SU(2)_R$ breaking (see also \cite{Diakonov:1984vw}).
In this case, it is natural to assume that
at $T=T_c$, the instantons disappear, reducing 
the $U(1)_A$ breaking effect, which makes 
the $SU(2)_L\times SU(2)_R$ to be recovered.

It has been a difficult issue in QCD to confirm
either of the two scinarios near $T=T_c$.
In analytic studies, it was found that the
semiclassical QCD instanton configurations 
are not enough to quantitatively describe the 
low-energy dynamics of QCD.
In lattice QCD simulations, it was difficult 
to treat chiral symmetries in a theoretically 
controled manner.
The $SU(2)_L\times SU(2)_R\times U(1)_A$
symmetry is explicitly broken down to 
an axial $U(1)'$ subgroup (different from $U(1)_A$)
symmetry in the staggered fermion formulation 
or down to a vectorlike $SU(2)$ symmetry 
in the Wilson fermon formulation.

In this work, we study the
chiral condensate and its susceptibility 
in 2- and $2+1$-flavor QCD 
with chiral symmetric Dirac operator on a lattice \cite{Neuberger:1997fp}.
We separate the $U(1)_A$ breaking 
effect or topological effect in particular,
from others in a theoretically clean way \cite{Aoki:2012yj, Cossu:2015kfa, Tomiya:2016jwr, Aoki:2020noz}.


\section{$U(1)_A$ contribution to chiral susceptibility}

In this work, we formally rewrite the quark determinant
in the QCD partition function
by the eigenvalues $\lambda$ of the Dirac operator.
Then the chiral condensate is a configuration average of
the summation over eigenvalues
\begin{align}
  \Sigma(m,T)
  = \frac{1}{N_fV}\frac{\partial}{\partial m}\ln Z(m)
  &=\frac{1}{V}\left\langle \sum_\lambda \frac{1}{i\lambda(A)+m}\right\rangle,
\end{align}
and the chiral susceptibility is decomposed into the connected part given by
the derivative with respect to the valence mass, and the disconnected part expressed by the derivative with respect to the sea quark mass:
\begin{align}
  \chi^\text{con.}(m)  &= \left.\frac{\partial}{\partial m_\text{valence}},
  \Sigma(m,T)\right|_{m_\text{valence}=m}\\
  \chi^\text{dis.}(m) &= \left.\frac{\partial}{\partial m_\text{sea}}
  \Sigma(m,T)
  \right|_{m_\text{sea}=m}.
\end{align}

We relate the chiral susceptibility to a set of the other susceptibilites
made of the scalar and pseudoscalar singlet operators : $S^0(x) = \bar{q}q(x)$
and $P^0(x) = \bar{q}i\gamma_5q(x)$ where $q(x)=(u,d)^T$ denotes the up and down
isospin doublet and those triplets: $S^a(x) = \bar{q}\tau^a q(x)$
and $P^a(x) = \bar{q}i\gamma_5\tau^a q(x)$, where $\tau^a$ denotes
the $a$-th isospin generator or the Pauli matrix.
These operators are connected by the $SU(2)_L\times SU(2)_R$
transformation $\exp(i\pi \gamma_5 \tau^a/2)$ and
the $U(1)_A$ one $\exp(i\pi \gamma_5/2)$.

By definition, the chiral susceptibility is equal to the
singlet scalar susceptibility
\begin{align}
 \chi(m)= - \sum_x \langle S^0(x)S^0(0)\rangle -V\langle S^0\rangle^2.
\end{align}
Adding the $\theta$ term and absorbing it into
the pseudoscalar mass term by the $U(1)_A$ rotation,
we can relate the topological susceptibility to
the pseusoscalar correlators,
\begin{align}
\frac{N_f}{m^2}\chi_{\rm top.}(m)
 = - \sum_x \langle P^0(x)P^0(0)\rangle +\frac{\Sigma(m,T)}{m}. 
\end{align}
From the Ward-Takahashi identity about the $SU(2)_L\times SU(2)_R$
transformation of the $P^a(x)$, we have
\begin{align}
m \sum_x \langle P^a(x)  P^a(0)\rangle + \langle S^0\rangle =0. 
\end{align}

Recombining the above identities we obtain
\begin{align}
\chi^{\rm con.}(m) &=\textcolor{black}{\underbrace{-\Delta_{U(1)}(m)+\frac{\langle |Q(A)|\rangle}{m^2V}}_\text{$U(1)_A$ breaking}
  \color{black}{\underbrace{+\frac{\Sigma_{\rm sub.}(m,T)}{m}}_\text{mixed.}}},\\
\chi^{\rm dis.}(m) &=\textcolor{black}{\underbrace{\frac{N_f}{m^2}\chi_{\rm top.}(m)}_\text{$U(1)_A$ breaking}} \textcolor{black}{\underbrace{+\Delta_{SU(2)}^{(1)}(m)-\Delta_{SU(2)}^{(2)}(m)}_\text{$SU(2)_L\times SU(2)_R$ breaking}},
\end{align}
where $Q(A)$ denotes the topological charge or the index
of the Dirac operator. As indicated in the equations,
the axial $U(1)$ susceptibility
\begin{align}
\Delta_{U(1)}(m) \equiv  
  \sum_x \langle P^a(x)P^a(0)-S^a(x)S^a(0)\rangle,
\end{align}
measures a purely $U(1)_A$ breaking effect,
while the other two susceptibilities
\begin{align}
\Delta_{SU(2)}^{(1)}(m) \equiv  
\sum_x \langle S^0(x)S^0(0)-P^a(x)P^a(0)\rangle,\;\;\;
\Delta_{SU(2)}^{(2)}(m) \equiv  
  \sum_x \langle S^a(x)S^a(0)-P^0(x)P^0(0)\rangle,
\end{align}
are the probes for the $SU(2)_L\times SU(2)_R$ breaking.
Note that we have subtracted the chiral condensate at
a reference quark mass $m_\text{ref}=0.005$ in order
to cancel the quadratic UV divergence, which is denoted by
\begin{align}
  \frac{\Sigma_{\rm sub.}(m,T)}{m}=\left[\frac{\Sigma(m,T)}{m}
    -\frac{\langle|Q(A)|\rangle}{m^2V}\right]-
  \left[\frac{\Sigma(m_\text{ref},T)}{m_\text{ref}}
    -\frac{\langle|Q(A)|\rangle|_{m=m_\text{ref}}}{m_\text{ref}^2V}\right].
\end{align}

Note in the discussion above we have used the notation in the continuum theory
but the corresponding lattice expression
can be given by the eigenvalues of the
massive overlap Dirac operator $H_m=\gamma_5[(1-m)D_\text{ov}+m]$ \cite{Aoki:2012yj}.
The precise chiral symmetry is, however, essential in the decomposition\cite{Nicola:2018vug}. 
Our goal of this work is to quantify how much the
$U(1)_A$ breaking contributes to the signal of the chiral susceptibility
in lattice QCD with exactly chiral symmetric Dirac operator.

\section{Numerical results}

We simulate $N_f=2$ QCD with a lattice spacing fixed at $1/a=2.6$ GeV.
We employ the Symanzik gauge action with $\beta=4.30$
and M\"obius domain-wall fermions \cite{Brower:2005qw}
for configuration generations
with the residual mass less than 1 MeV. The quark mass is taken in a range $[3,30]$ MeV, which
covers a point below the physical up and down quark mass
and we have six different temperatures obtained by changing the temporal lattice size $L_t$
from 8 to 18, which corresponds to $[147\text{--}330]$ MeV. 
Our main lattice size is 32 or $2.4$ fm but for lower temperatures
we extend to 40($\sim 3$fm)  or 48($\sim 3.6$fm) to study the finite volume systematics.
As will be discussed below, the pseudo-critical temperature is estimated to be $T_c\sim 165$ MeV
(at the physical point)
and the lowest simulated temperature on $L_t=18$(147 MeV) lattices correspond to $0.9T_c$.

In order to reduce the lattice artifact due to the violation  
of the Ginsparg-Wilson relation \cite{Cossu:2015kfa}, 
we use the reweighting technique
to replace the M\"obius domain-wall fermion determinant by
that of the overlap fermion action, in which the sign function
for the low modes of the kernel Dirac operator are taken exactly.
The residual mass is reduced to 0.001 MeV with this procedure.
We also increase the simulation points by the same reweighting
technique but with different quark mass.
With the mass reweighting, we obtain the results down to
$m=0.0002$ or 1/5 physical point mass.

In this work, the observables are computed with the spectral decomposition 
or the summation over the eigenvalues.
The cutoff of the summation at 30--40-th lowest modes,
which corresponds to 150--300 MeV, gives a  good saturation and consistency with direct inversion of M\"obius domain-wall Dirac operator as far as $T\le 260$~MeV.
At $T=$330 MeV, the convergence is marginal and we do not employ the low-mode approximation.

\begin{figure*}[bth]
  \centering
  \includegraphics[width=10.5cm]{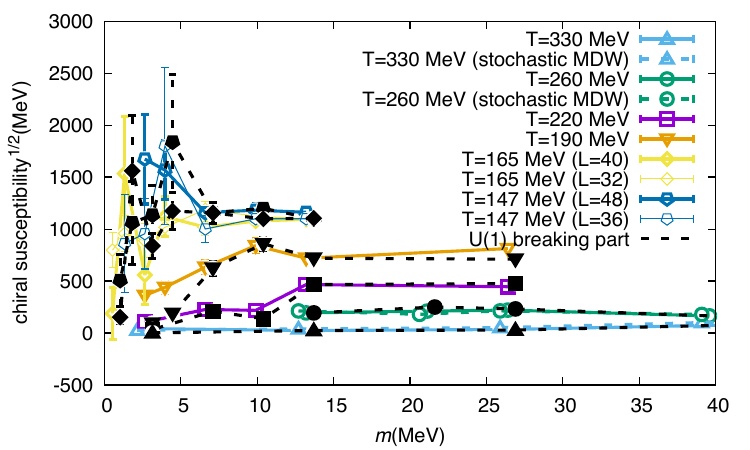}
  \includegraphics[width=10.5cm]{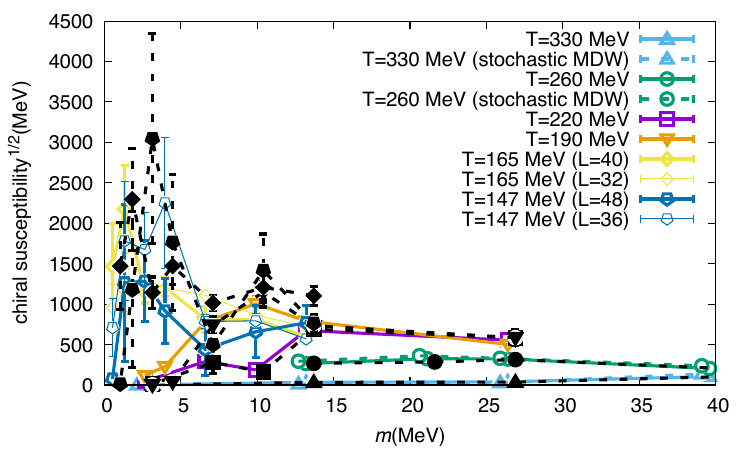}
  \caption{
    The connected (top panel) and disconnected (bottom) susceptibilities 
    are plotted as a function of the quark mass at 6 different temperatures.
    Compared to the total contribution in open solid symbols,
the axial $U(1)$ anomaly part in dashed filled are dominant.
But this $U(1)_A$  dominance becomes unclear and noisier for light quark mass 
points at lower temperatures.
  }
  \label{fig:result-summary}
\end{figure*}

\begin{figure*}[bth]
  \centering
  \includegraphics[width=10.5cm]{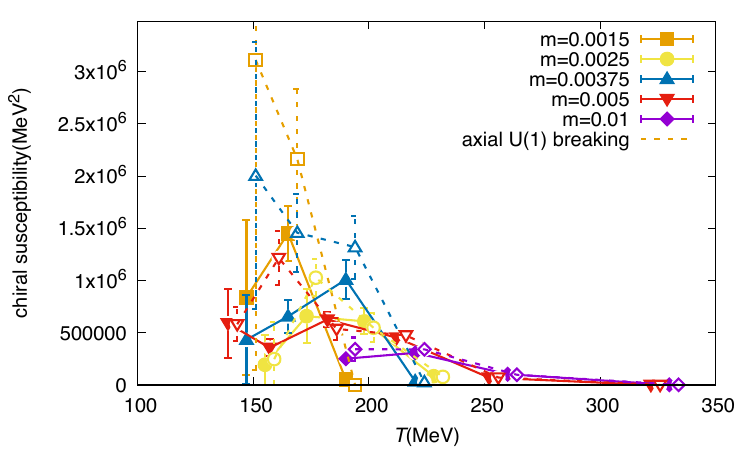}
  \includegraphics[width=10.5cm]{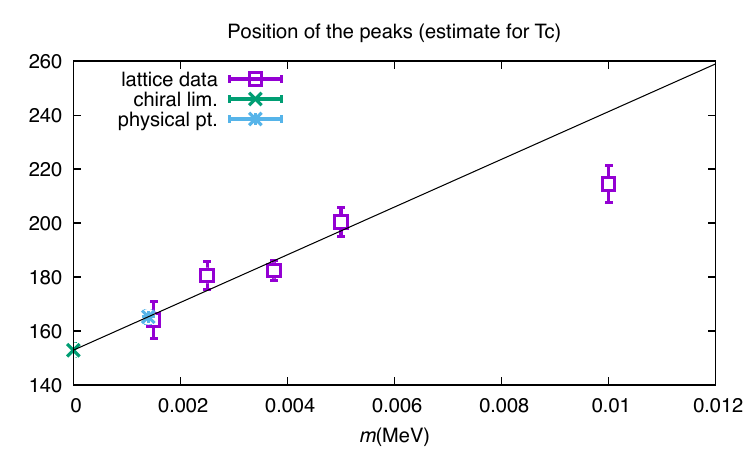}
  \caption{
    Top: $T$ dependence of the  disconnected susceptibility
    at 5 different quark masses.
 Bottom: The chiral extrapolation of the peak position
of the disconnected susceptibility gives an estimate for the
(pseudo-) critical temperature. Bottom: Our estimate for $T_c$ from the
quark mass dependence of the peak position of the disconnected susceptibility.
  }
  \label{fig:result-summary2}
\end{figure*}

In Fig.~\ref{fig:result-summary} the connected (top panel) and disconnected (bottom) chiral susceptibilities 
    are plotted as a function of the quark mass at 6 simulated temperatures.
The open solid symbols are the lattice data, while the filled dashed symbols are $U(1)_A$ breaking part.
We can see that the $U(1)_A$ breaking contribution dominates the signals.
Specifically the axial $U(1)$ susceptiblity dominates the connected part of the chiral susceptibility
and the topological susceptibility divided by the quark mass squared
is respobsible for the signal of the disconnected chiral susceptibility. 
This is the case even for the lower temperature around 147 MeV in pentagons and lower quark masses 
obtained by the reweighting and larger volumes. 
However, this $U(1)_A$  dominance becomes less obvious with noisier signals for light quark mass 
points at lower temperatures.
This behavior is shown in the top panel of Fig.~\ref{fig:result-summary2}
where the $T$ dependence of the disconnected susceptibility is presented.
Fitting data to a quadratic function, we can estimate the position of the peak
at each quark mass.
The chiral extrapolation of the susceptibility peak is shown in
the bottom panel of Fig.~\ref{fig:result-summary2}.
The preliminary estimate for 
the (pseudo-) critical temperature is $T_c\sim 165$ MeV at the physical point,
and $T_c\sim 153$ MeV in the chiral limit.


\section{Summary}

We simulate $N_f=2$ QCD at high temperatures.
The chiral condensate and susceptibility are related not only to the standard
$SU(2)_L\times SU(2)_R$ chiral symmetry but also to the anomalous $U(1)_A$ symmetry.
In the spectral decomposition of the Dirac operator with exact chiral symmetry on a lattice, 
we can separate the purely $U(1)_A$ breaking effect.
We have found that the connected part is dominated by the axial $U(1)$ susceptibility,
and the disconnected part is governed by the topological susceptibility.
But for lower $T$, the deviation becomes sizable.
Our results suggest that the axial $U(1)$ anomaly may play more important role 
in the QCD phase diagram \cite{Pisarski:1983ms} than expected.

We thank L. Glozman and Y. Sumino for useful discussions. 
For the numerical simulation we have used the QCD software packages
Grid \cite{Boyle:2015tjk,Meyer:2021uoj} for configuration generations
and Bridge++ \cite{Ueda:2014rya,Akahoshi:2021gvk} for measurements.
Numerical simulations were performed on Oakforest-PACS and Wisteria/BDEC-01 
at JCAHPC under a support of the HPCI System Research Projects 
(Project IDs: hp170061, hp180061, hp190090, and hp200086, hp210104, hp220093, hp230070) as well as 
Multidisciplinary Cooperative Research Program in CCS, University of Tsukuba
(Project IDs: xg17i032, xg18i023 and wo22i038) and 
Fugaku computer provided by the RIKEN Center
for Computational Science under a support of the HPCI System Research Projects 
(Project IDs: hp200130, hp210165, hp210231, hp220279, hp230323).
This work was (partly) achieved through the use of SQUID at the Cybermedia Center, Osaka University.
We used Japan Lattice Data Grid (JLDG) \cite{Amagasa:2015zwb} for storing a
part of the numerical data generated for this work. This work is supported in part by the
Japanese Grant-in-Aid for Scientific Research (No. JP26247043, JP18H01216,
JP18H04484, JP18H05236, JP22H01219) and by Joint
Institute for Computational Fundamental Science (JICFuS).

\end{document}